\documentclass[dvipdfmx]{revtex4}
\usepackage{graphicx,color}
\usepackage{epsf,bm,amsmath}
\def\bk{\mbox{\boldmath $k$}}
\newcommand{\et}{\textit{et al. }}
\begin{document}

\title{Nuclear saturation in lowest-order Brueckner theory with two- and three-nucleon
forces in view of chiral effective field theory}
%\author{\name{M. Kohno}{1}}
%\address{\affil{1}{Research Center for Nuclear Physics, Osaka University, Ibaraki 567-0047, Japan}
%\email{kohno@rcnp.osaka-u.ac.jp}}
\author{M. Kohno\footnote{kohno@rcnp.osaka-u.ac.jp}}
\affiliation{Research Center for Nuclear Physics, Osaka University, Ibaraki 567-0047, Japan}

\begin{abstract}
The nuclear saturation mechanism is discussed in terms of two-nucleon and
three-nucleon interactions in chiral effective field theory (Ch-EFT), using the framework
of lowest-order Brueckner theory. After the Coester band, which is observed in
calculating saturation points with various nucleon-nucleon (NN) forces, is revisited
using modern NN potentials and their low-momentum equivalent interactions,
detailed account of the saturation curve of the Ch-EFT interaction is presented.
The three-nucleon force (3NF) is treated by reducing it to an effective two-body
interaction by folding the third nucleon degrees of freedom. Uncertainties due
to the choice of the 3NF low-energy constants $c_D$ and $c_E$ are discussed.
The reduction of the cutoff-energy dependence of the NN potential is explained
by demonstrating the effect of the 3NF in the $^1$S$_0$ and $^3$S$_1$ states.
% \textcolor{red}{\today}
\end{abstract}

%\subjectindex{D04}

\maketitle

\section{Introduction}
It is one of the fundamental problems in nuclear physics to understand saturation
properties of atomic nuclei on the basis of the underlying interactions between nucleons.
A description of the nucleon-nucleon (NN) interaction has been developed in about
eighty years after the meson theory was conjectured by Yukawa in 1935 \cite{YUK35}.
It was recognized in the early stage from the analyses of NN scattering data that
the NN force described as an instantaneous two-nucleon potential is strongly repulsive
at short distances as is fairly well approximated by a hard core. On the other hand,
experimental evidences that disclose single-particle structures of atomic nuclei
have been accumulated. An explanation of this seemingly contradicting situation was
given by Brueckner theory in the 1950's \cite{BLM54}. The central element in this
theory is a reaction matrix (or $G$ matrix), which describes in-medium NN correlation
and takes care of strong repulsion of the NN interaction at short distances.
Afterwards, the theory was organized as the quantum many-body theory
\cite{GOL57,DAY67,BET71}: the framework of a perturbation description of
nuclei in terms of the $G$-matrices.

Although the Brueckner theory was successful to qualitatively account for the reason
for the appearance of the shell structure, a quantitative explanation of nuclear
saturation properties has not been completed. Various calculations in nuclear matter
showed that theoretical saturation points obtained from existing NN potentials which
achieve very high accuracy in describing scattering data miss the empirical location
\cite{DAY78,LSZ06}. It is improbable that a certain new functional form of the potential,
either in coordinate space or in momentum space, resolves the problem. Concerning
higher-order correlations not included in these nuclear matter calculations, their
contributions have been shown to be rather modest and do not improve the situation \cite{SBGL98}.
In addition, different quantum many-body frameworks, for example a variational method,
also predict similar saturation properties to those in the Brueckner theory \cite{PW79,BB12}.

There have been various conjectures about the missing mechanism to explain the
correct nuclear saturation properties, such as relativistic effects, many-body
forces, and possible modifications of nucleon properties in the nuclear medium.
Among them, the three-nucleon force (3NF) contribution is a promising candidate to be
investigated first, for which the Fujita-Miyazawa \cite{FM57} type interaction
involving an isobar $\Delta$ excitation is a prototype. As for the relativistic effect, Dirac-Bruckner
calculations were shown to explain quantitatively nuclear matter saturation \cite{BM90}.
However, the processes including anti-nucleon excitations involve considerably higher
energy scale than that of the isobar $\Delta$ excitation. Thus, it is important to establish
the contribution of the three-nucleon force effects to examine if it is reasonable or not.
There have been many attempts in the literature to use the 3NF contributions to reproduce
correct nuclear saturation properties \cite{BB12,BG69,LNR71,KAT74,FP81}. However,
these calculations are exploratory or include more or less phenomenological adjustments.

The situation has changed since the description of the NN interaction
in chiral effective field theory (Ch-EFT) \cite{EHM09,ME11} achieved,
at the N$^3$LO level, the comparable accuracy with other modern NN interactions
below the pion production threshold energy
and the 3NFs were introduced systematically and consistently to the NN sector. 
The study of Ch-EFT 3NF contributions in nuclear and neutron matter started
in the many-body framework such as the Hartree-Fock approximation
\cite{BSFN05}, and perturbation calculations up to the second order \cite{HS10}
and the third order \cite{HEB11,CORA14}.

The present author gave a brief report, in Ref. \cite{MK12}, on the
lowest-order Brueckner theory (LOBT) calculations in nuclear matter
using the N$^3$LO NN and NNLO 3NF interactions, in which a focus was put on
the effective spin-orbit strength. The 3NF was treated as a
density-dependent effective NN force, as in Holt \et \cite{HKW10} but without introducing
an approximation for  off-diagonal matrix elements. More details about the
calculated saturation curves were presented in Ref. \cite{MK13}.
Sammarruca \et \cite{SCC12} presented results of
similar LOBT calculations. They also discuss cutoff scale dependence and
order-by-order convergence of the Ch-EFT interactions on nuclear and neutron
matter calculations \cite{SCC15} . Carbone \et \cite{,CARB13,CARB14}
have investigated the Ch-EFT 3NF contributions in their
self-consistent Green's function formalism, taking into account a
correlated average of the 3NF over the third nucleon.
All these calculations find the important and desirable effects of the 3NF to
improve the description of nuclear saturation properties.

It was shown in Ref. \cite{MK13} that the saturation curve becomes
close to the empirical one when the effects of the 3NF are included as an
effective NN interaction by folding the third nucleon. In addition, the strong
spin-orbit strength required for nuclear mean filed calculations is explained
by the additional contribution from the 3NF \cite{MK12,MK13}.
A further interesting observation is that the cutoff-energy dependence was
substantially reduced by the inclusion of the 3NF effects, which
is a desirable feature for the effective theory
having a cutoff parameter. The mechanism to produce these results was briefly
discussed in Ref. \cite{MK13}. This paper supplements the explanation by extending
the discussion to the wide scope of the nuclear saturation mechanism.

In fermion many-body calculations, a Pauli exclusion plays essential roles in
considering two-particle correlations. In a nucleon-nucleon correlation in the
nuclear medium, the Pauli effect reveals dominantly in the triplet even channel,
because typical momentum transfer in the tensor correlation lies in the momentum
region of the Fermi momentum of nuclei. On the other hand, short-range repulsion
is characterized by higher momentum transfer, and hence the short-range correlation
is not sensitive to the Pauli effect. The introduction of equivalent interactions
in low-momentum space \cite{BKS03} is concerned with the elimination of the
short-range and thus high-momentum component of the NN interaction in the two-nucleon
sector. It is useful to consider nuclear matter saturation curves obtained
using low-momentum equivalent interaction. Because the tensor interaction is also
transformed to fit in the low-momentum space, the saturation curve varies
with changing the low-momentum scale.
This variation is due to the elimination of possible correlations in the many-body space,
which should be recovered by inducing many-body effective interactions in the
restricted space. This situation is shown to be analogous to the 3NF effects.

Section 2 revisits a Coester band of LOBT nuclear matter calculations, which was first
demonstrated by Coester \et \cite{COE70} in 1970. The character of unitary
uncertainties in the NN potential on the LOBT calculation is further demonstrated
in Sect. 3 by using low-momentum interaction, which is equivalent to the original
interaction in the low-momentum two-body space. Section 4 reports the results of the
nuclear matter calculations by using Ch-EFT interactions. First, uncertainties due
to the 3NF parameters $c_D$ and $c_E$ are discussed  in subsection 4.1. It is shown
that the $c_D$ and $c_E$ term contributions to nuclear matter energies cancel each other
when $c_D\approx 4c_E$ is satisfied, and this relation is favorable to describe
nuclear saturation properties as shown in Ref. \cite{MK13}. On the basis of this observation,
the results with $c_D=c_E=0$ are first presented and the dependence of calculated energies
on $c_D$ and $c_E$ is demonstrated. Qualitative explanations of the 3NF contributions
are given in Section 5. Summary and some remarks follow in Sec. 6.

\section{Coester band of LOBT saturation points}
The Coester band of LOBT nuclear matter calculations is recapitulated
in order to give the basis for the discussion in the following sections.

In the lowest-order of the Brueckner-Bethe-Goldstone many-body theory, the energy per
nucleon in nuclear matter with the Fermi momentum $k_F$ is given by
\begin{equation}
  E/A=\frac{1}{\rho} \sum_{|\bk|\le k_F} \frac{\hbar^2}{2m}k^2
   + \frac{1}{2\rho} \sum_{|\bk_1|,|\bk_2| \le k_F} \langle \bk_1\bk_2|G|\bk_1\bk_2\rangle_A,
\end{equation}
where spin and isospin summations are implicit and the subscript $A$ stands for the
antisymmetrization. The nucleon density of symmetric nuclear matter is calculated
as $\rho=\sum_{|\bk|\le k_F} 1=\frac{2k_F^3}{3\pi^2}$.
The reaction matrix $G$ is determined by the $G$-matrix equation
%\begin{widetext}
\begin{align}
  & G|\bk_1\bk_2\rangle_A  =   V|\bk_1\bk_2\rangle_A \notag \\
  & \hspace{1em}+ \sum_{\bk_1'\bk_2'}V|\bk_1'\bk_2'\rangle_A
 \langle \bk_1'\bk_2'| \frac{Q}{e(k_1)+e(k_2) -e(k_1')-e(k_2')}|\bk_1'\bk_2'\rangle_A
 \langle \bk_1'\bk_2'|G|\bk_1\bk_2\rangle_A,
\end{align}
%\end{widetext}
where $V$ denotes a two-nucleon bare interaction.
The single-particle energy $e(k)$ of an occupied state, $k \le k_F$, is determined
from the $G$-matrix self-consistently:
\begin{equation}
 e(k)= \frac{\hbar^2}{2m}k^2+\sum_{|\bk'| \le k_F} \langle \bk\bk'|G|\bk\bk'\rangle_A.
\end{equation}
This definition corresponds to including diagrams shown in Fig. 1 in the energy.
As for the unoccupied state, there are some ambiguities linked to the choice of
intermediate spectra of the perturbative expansion. There are two standard prescriptions.
The gap choice takes $e(k)=\frac{\hbar^2}{2m}k^2$ for $k > k_F$.
The reason for this prescription is that the insertion for the
particle state should be treated as the three-body correlation diagrams and the net
contribution of these diagrams is small. The other is the continuous choice; that is,
$e(k)=\frac{\hbar^2}{2m}k^2+\sum_{|\bk'| \le k_F} \langle \bk\bk'|G|\bk\bk'\rangle_A$
for $k > k_F$. In this case, there is no rigorous correspondence similar to that
in Fig. 1, because $G$-matrix cannot be on-shell for the insertion in the perturbative
expansion of the total energy. Since Song \et \cite{SBGL98} showed that the net
contribution in the total energy from
higher-order diagrams is smaller in the continuous prescription, this choice
has been commonly employed.
In the calculations below, the Pauli exclusion operator is approximated by
introducing angle-averaging. The accuracy of this approximation
in Ref. \cite{SMC99,SOKN00} for the gap prescription. The denominator of Eq. (2)
is also treated with an angle-average for $\bk_1'$ and $\bk_2'$.

\begin{figure}[t]
\centering
\includegraphics[width=0.5\textwidth]{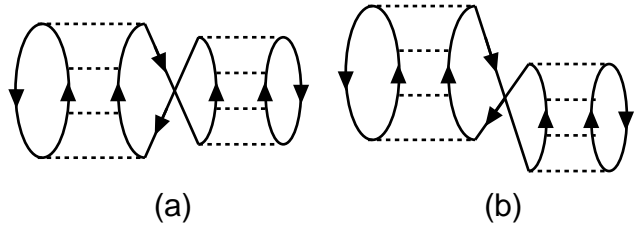}%fig1.eps=hole-ins.eps
\caption{Three-hole line diagrams corresponding to the definition
of single-particle potential, Eq. (3). The downwards (upwards) arrow
represents an occupied (unoccupied) state. The dotted line denotes a
bare interaction $V$.
}
\end{figure}

Nuclear matter calculations in the LOBT have shown \cite{DAY78,LSZ06} that varying
nucleon-nucleon potentials predict different saturation curves; namely,
the relation of the calculated energy per nucleon $E/A$ to the Fermi momentum of
nuclear matter $k_F$. The salient feature is that the saturation point of
the different interaction,
the minimum of the saturation-energy curve, lies in a rather narrow band, in which
the empirical saturation point does not locate. This band has been called as the
Coester band, after Coester \et presented such systematic results in Ref. \cite{COE70}.
This property persists in the calculations using modern nucleon-nucleon potentials with
higher accuracy in reproducing NN scattering data. Various calculations in the literature
indicate that higher-order correlations beyond the standard Brueckner calculation do
not change the situation, although a different choice of the intermediate spectra of the
propagator in the $G$-matrix equation causes a shift of the Coester band itself. Figure 2
demonstrates saturation curves obtained in the LOBT with the continuous (thick curves) and
gap (thin curves) prescriptions for the intermediate spectra of the propagator
in the $G$-matrix equation, using NN potentials such as AV18 \cite{AV18},
NSC97 \cite{NSC}, CD-Bonn \cite{CDB}, and fss2 \cite{FSS2}. 
  
\begin{table}[b]
 \caption{Deuteron $D$-state probability $p_D$ of the NN interactions used in Fig. 2.}
 \setlength{\tabcolsep}{12pt}
 \begin{center}
 \begin{tabular}{ccccc} \hline\hline
   & AV18 & NSC97 & fss2 & CD-Bonn \\ \hline
 $p_D$ & 5.76 & 5.39 & $5.49$ & 4.85 \\ \hline\hline
 \end{tabular}
 \end{center}
\end{table}

The variation of the saturation curve is known to be controlled by the
strength of the tensor component. Because the considerable attraction in the
proton-neutron ($^3$S$_1$) channel is brought about by ladder correlation through
the tensor force and the correlation is influenced by the Pauli effect in
the nuclear medium, the binding energy of nuclear matter depends on
the weight of the tensor component of the NN potential used. A useful measure
to specify the strength of the tensor correlation is a deuteron $D$-state
probability $p_D$, although this quantity is not measurable.
Table 1 tabulates values of $p_D$ of the NN interactions used in Fig. 2.

The NN potential with larger $p_D$, which provides attractive contribution by the
tensor correlation in free space, tends to give a shallower saturation curve
in nuclear matter due to the suppression of the tensor correlation.
Although this qualitative expectation does not explain why the shift due to the
difference of $p_D$ moves in the narrow Coester band, it is apparent that we
should expect the variation of the saturation point depending on the NN
potential used in many-body calculations.

\begin{figure}[t]
\centering
\includegraphics[width=0.5\textwidth]{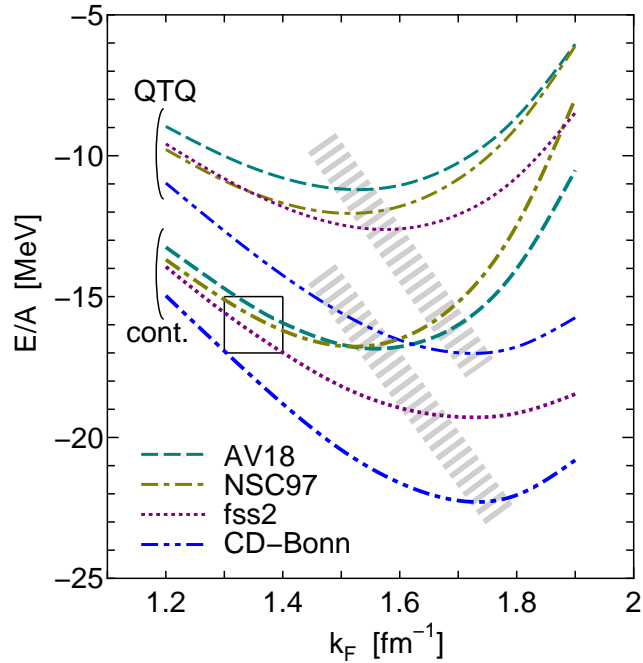}%fig2.eps=modern-nmbe.eps
\caption{LOBT saturation curves in symmetric nuclear matter with the continuous (thick curves)
and the gap (thin curves) prescriptions for the intermediate spectra,
using modern NN potentials: AV18 \cite{AV18}, NSC97 \cite{NSC},
CD-Bonn \cite{CDB}, and fss2 \cite{FSS2}. Two broad hatches are a rough guide for
the Coester band. The standard empirical saturation point is indicated by the square.
}
\end{figure}

It is worthwhile to note that most of saturation curves calculated with the
continuous prescription go past near the
empirical saturation point which is indicated by the square in Fig. 2. It implies that
the $G$-matrices as an effective interaction quantitatively work well for describing
various structural and scattering properties of nuclei, as far as the quantities
associated with the slope and/or the curvature of the saturation curve are not
crucial for them.

\section{LOBT Calculations using low-momentum interactions}
A different approach to manipulate the singular short-range part of the NN force
was developed in the 1990's: that is, low-momentum interaction theory \cite{BKS03}.
The method is based on a renormalization group viewpoint and equivalent
interaction theory in a restricted space. Starting from the bare NN interaction,
high-momentum components are integrated out to define the interaction $V_{low k}$ in
low-momentum space appropriate for low-energy nuclear physics. $V_{low k}$ is the
interaction which reproduce the same half-on-shell $T$ matrices of the original bare
NN interaction $V$. Denoting the projection operator into the low-momentum space by $P$,
$V_{low k}$ is defined to satisfy
\begin{equation}
 PTP= PV_{low k}P+PV_{low k}P\frac{1}{\omega-t}PTP,
\end{equation}
where $\frac{1}{\omega -t}$ is a free nucleon propagator and $T$-matrix is given
by $T=V+V\frac{1}{\omega-t}T$ in the entire space.

It is straightforward to do nuclear matter LOBT calculations by taking
the low-momentum interaction as the input NN interaction. Because high-momentum
components are eliminated, the $G$-matrix equation is not meant for taking care of
them, but the $G$-matrix equation and the self-consistency of single-particle energies
take into account ladder correlations together with certain higher order correlations
in the low-momentum space. By definition, low-momentum interaction is
regarded as obtained by a unitary transformation, because low-momentum
interaction in the Lee-Suzuki method \cite{SL80,LS80} is identical to that of
the renormalization group consideration \cite{BKS03}. In this sense,
the low-momentum interaction is a soft version of the original bare interaction
by some appropriate unitary transformation. Thus we expect that the saturation
point obtained by $V_{low k}$ with a different low-momentum scale moves in the Coester band.
As an illustration, the AV18 \cite{AV18} and CD-Bonn \cite{CDB}
potentials \cite{AV18,CDB} are taken as a starting bare potential and construct low-momentum
interaction for the cutoff of $\Lambda_{low k}=6$, $5$, $4$, and $3$ fm$^{-1}$, respectively.
The saturation curves obtained in the LOBT with these potentials are plotted in Fig. 3.
Varying the low-momentum cutoff $\Lambda_{low k}$, the saturation point systematically
shifts on the Coester band. The similar result was presented before by
Kuckei \et \cite{KMMS03}. The difference of the saturation curves
obtained from the AV18 and CD-Bonn is seen to gradually reduce by lowering the
low-momentum scale $\Lambda_{low k}$.

\begin{figure}[t]
\centering
\includegraphics[width=0.5\textwidth]{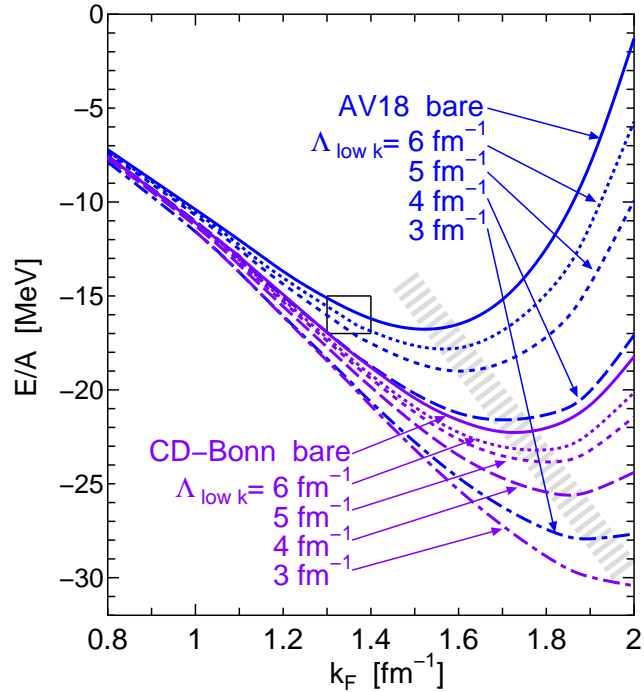}%fig3.eps=lowk-sole.eps
\caption{Variation of LOBT saturation curves in symmetric nuclear matter
under the change of a low-momentum cutoff scale $\Lambda_{low k}$. Results for
two NN potentials, AV18 \cite{AV18} and CD-Bonn \cite{CDB}, are shown. The broad hatch
is a rough guide for the band in which saturation minima locate.
}
\end{figure}

The unitary transformation of the NN interaction preserves corresponding
on-shell properties of the original interaction in the low-momentum space.
However, it induces
many-body interactions when practiced in a many-body space. If these
induced many-body interactions are included, the result of the original bare
interaction should be recovered.
The occurrence of this restoration was actually demonstrated in few-body
calculations in the similarity renormalization group method \cite{JNF11}.
In the case of low-momentum interaction method, the unitary transformation
to the low-momentum space is to be carried out in the many-body space.
Such a framework to include the induced interaction was developed by Suzuki,
Okamoto, and Kumagai \cite{SOK87} as the unitary-model-operator-method
(UMOA), and was applied for $^{16}$O and $^{40}$Ca in \cite{FOS09}.
However, an explicit application of this method in nuclear matter has not been
undertaken. The inclusion of higher-order correlations are
beyond the scope of the present paper.

The necessity of  incorporating the induced many-body interaction to recover the
result of the original bare force is analogous to the introduction of the 3NF and the
cutoff-scale dependence of its contributions. This point is discussed in the following
sections.

\section{LOBT calculations with the N$^3$LO NN interactions in chiral effective
field theory}
This section supplements the results of Ref. \cite{MK13}, where LOBT nuclear matter
calculations are presented, using NN and 3N forces in chiral effective field theory.
The NNLO 3NF has 5 low-energy constants. Three of them are fixed in the NN sector:
 $c_1=-0.81$ GeV$^{-1}$, $c_3=-3.4$ GeV$^{-1}$, and $c_4=3.4$ GeV$^{-1}$ \cite{EGM05}.
Remaining parameters, $c_D$ and $c_E$, are adjusted to reproduce, for example,
observables of few-nucleon systems. In Ref. \cite{MK13}, the values of $c_D=-4.381$
and $c_E=-1.126$ given in the Table 1 of the paper by Hebeler \textit{et al.} \cite{HEB11}
for the EGM interaction \cite{EGM05} were employed, and uncertainties due
to $c_D$ and $c_E$ were disregarded because the nuclear matter saturation curves
are reasonably well reproduced. In this paper, a more careful analysis is prepared for
the uncertainties on these constants. Using a Hartree-Fock level expression, it is
shown that if the relation of $c_D \approx 4c_E$ is satisfied,
contributions of these two terms almost cancel each other.
The values used in Ref. \cite{MK13} approximately fulfil this relation. It indicates
that the condition of $c_D \approx 4c_E$ is preferable for the LOBT calculations
in nuclear matter. It is interesting that
the few-nucleon calculations \cite{NGV07} actually suggests $c_D \approx 4c_E$ in
the case of $c_D < 0$ in its continuous uncertainties. Therefore, in this paper,
calculated results with $c_D =c_E=0$ are first presented as a reference case and
the variation due to changing  $c_D$ and $c_E$ is shown.

\subsection{Uncertainties of $c_D$ and $c_E$}
%The 3NF defined at the NNLO level of chiral effective field theory involves three coupling
%constants $c_1$, $c_3$ and $c_4$ which are determined in the 2NF sector and two
%new coupling constants $c_D$ and $c_E$. The last two coupling constants have to
%be adjusted to reproduce experimental data of more than three-nucleon systems. The analyses by
%Navratil \et \cite{NGV07} show that the binding energies of the $^3$H and $^3$He
%constrain the relation between $c_D$ and $c_E$ and \textit{ab initio} calculations
%of $p$-shell nuclei suggest the preferred values of $c_D\approx -1$ and $c_E\approx -0.35$.
%On the other hand, Noga \et \cite{NBS04} use $c_D=-2.06$ and $c_E=-0.63$ in their
%few-body calculations. The perturbative calculations of nuclear matter energies by
%Hebeler \et \cite{HEB11} use $c_D=-4.381$ and $c_E=-1.126$. These calculations
%indicate that the relation $c_D \approx (3\sim 4) c_E$ is appropriate to describe empirical
%nuclear properties, though it is to be kept in mind that the explicit values depend
%on the cutoff scale and calculating scheme.

Before carrying out explicit $G$-matrix calculations in nuclear matter, it is instructive to
consider the mean-field contributions of the $c_D$ and $c_E$ terms to the nuclear
matter energy. Integrating three-nucleon matrix elements in symmetric nuclear
matter, we obtain the following expression for the contributions to the energy, which
are same as those given by Bogner \et in Ref. \cite{BSFN05}.
\begin{align}
 E_{c_D}/A = & -\frac{3^5}{4(2\pi)^4} \frac{1}{k_F^3}
 \frac{g_A c_D}{f_\pi^4\Lambda_{\chi}} \int_0^{k_F} Y^2dY  \int_0^{\sqrt{k_F^2-Y^2}} p^2dp
 \int_0^{\frac{2}{3}(k_F+Y)} q^2dq  \notag \\
 & \times \frac{4p^2}{4p^2+m_\pi^2}f_R^2(p,q) F(p,Y;k_F)G(q,Y;k_F), \\
 E_{c_E}/A  =& -\frac{3^5}{(2\pi)^4}\frac{1}{k_F^3}\frac{c_E}{f_\pi^4\Lambda_\chi}
 \int_0^{k_F} Y^2dY \int_0^{\sqrt{k_F^2-Y^2}} p^2dp  \int_0^{\frac{2}{3}(k_F+Y)}q^2dq \notag \\
 & \times f_R^2(p,q) F(p,Y;k_F)G(q,Y;k_F),
\end{align}
where $f_R=\exp \{-(p^2+\frac{3}{4}q^2)^2/\Lambda_{3NF}^4\}$ is a regularization factor and
two functions $F(p,Y;k_F)$ and $G(q,Y;k_F)$ are defined as
\begin{align}
 F(p,Y;k_F)&=\left\{ \begin{matrix}   2 & \mbox{for $p\leq k_F-Y$},  \\
                                          \frac{k_F^2 -p^2-Y^2}{Yp} & \mbox{for $k_F-Y\leq p\leq \sqrt{k_F^2-Y^2}$}.
                              \end{matrix} \right. \\
 G(q,Y;k_F)&= \left\{ \begin{matrix} 2 & \mbox{for $0\leq q\leq \frac{2}{3}(k_F-Y)$}, \\
                                  \frac{k_F^2 -(\frac{3}{2}q-Y)^2}{3Yq} &
  \mbox{for $\frac{2}{3}(k_F-Y)\leq q\leq \frac{2}{3}(k_F+Y)$}.
                             \end{matrix} \right.
\end{align}
Numerical calculations in the range of $0.8 \leq k_F \leq 1.8$ fm$^{-1}$ with $f_\pi =92.4$ MeV,
$\Lambda_\chi=700$ MeV, and $\Lambda_{3NF}=2$ fm$^{-1}$
show that the values of $E_{c_D}/A$ and $E_{c_E}/A$ is fitted well
by simple quadratic polynomials of the function of the density
$\rho = \frac{2k_F^3}{3\pi^2}$ as
\begin{eqnarray}
\frac{E_{c_D}(\rho)}{A} &=& c_D\times (-0.1902 + 2.952 \rho + 37.16 \rho^2), \\
\frac{E_{c_E}(\rho)}{A} &=& c_E\times (0.8695 - 17.52\rho - 128.3\rho^2).
\end{eqnarray}
It is easy to see that $E_{c_D}/A(\rho)$ and $E_{c_E}/A(\rho)$ cancel to a large
extent at any density of $0.8 \leq k_F \leq 1.8$ fm$^{-1}$, namely $0.035 \leq \rho \leq
0.394$ fm$^{-3}$, when $c_D \approx 4 c_E$ holds. As is shown below, this cancellation
persists in the LOBT calculations.
The calculations in nuclear matter in Ref. \cite{MK13} indicate that contributions
from the $c_D$ and $c_E$ terms are better to cancel to describe nuclear properties.
As far as the nuclear matter energies are concerned, the actual values
of $c_D$ and $c_E$ are not important if $c_D \approx 4c_E$. It is even acceptable
to set $c_D=c_E=0$.
% which was explicitly tested in nuclear matter calculations.
Therefore, the main 3NF contributions come from the terms of the coupling
constants $c_1$, $c_3$ and $c_4$ which are settled in the NN sector, though
there is a possibility of carefully tuning $c_D$ and $c_E$ in a natural size to achieve a good
description of finite nuclei.
%In the calculations below, $c_D=-4.381$ and $c_E=-1.126$
%are used together with the preset coupling constants $c_1=-0.81$ GeV$^{-1}$,
%$c_3=-3.4$ GeV$^{-1}$, and $c_1=3.4$ GeV$^{-1}$ \cite{EGM05}.

\subsection{Results of LOBT calculations}
Figure 4 shows saturation curves obtained by the N$^3$LO Ch-EFT NN interaction
with the three choices of the cutoff scale: $\Lambda=450$, $550$, and $600$ MeV,
respectively. The results of the AV18 force \cite{AV18} shown in Fig. 3 are also
included for comparison. As low-energy effective
theory, the Ch-EFT potential is not applied to the high momentum region.
Hence, the LOBT calculation in Fig. 4 is limited below $k_F=1.8$ fm$^{-1}$.
The Ch-EFT potential with $\Lambda=550$ MeV is seen to possess a similar saturation
property in LOBT to that of AV18. Larger (smaller) cutoff energy affords
stronger (weaker) tensor components, and the saturation minimum obeys the
Coester band.

\begin{figure}[b]
\centering
\includegraphics[width=0.5\textwidth]{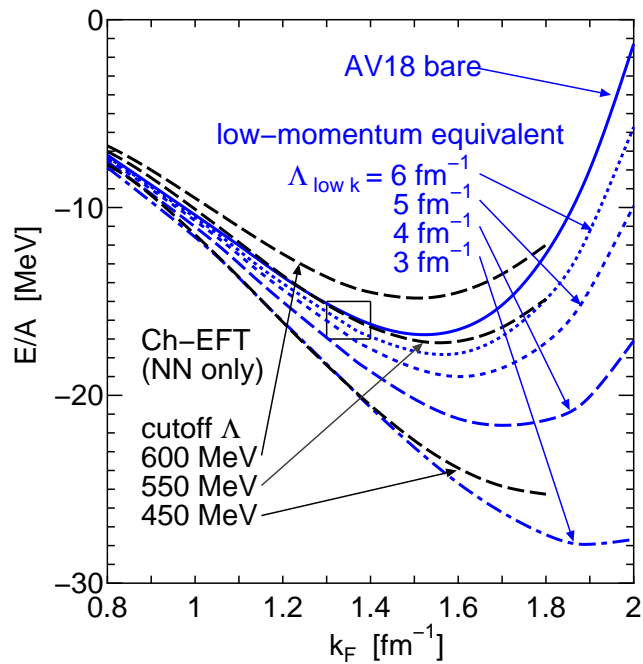}%fig4.eps=av18-lowk14.eps
\caption{LOBT saturation curves in symmetric nuclear matter with the continuous
choice for intermediate spectra,
using the N$^3$LO NN interaction with three choices of the cutoff energy:
$\Lambda=450$, $550$, and $600$ MeV, respectively. The results of the AV18
force \cite{AV18} shown in Fig.3 are also plotted for comparison.
}
\end{figure}

When the cutoff scale is
small, $\Lambda=450$ MeV, the tensor component is relatively small and thus
the central component gives larger attraction which is not much suppressed
by the Pauli blocking. On the other hand, if the attraction provided by the tensor
correlation is large in the free space, the attraction is sensitive to the Pauli effect.
This explains that the saturation point appears at the lower density with the smaller
binding energy. If $E/A$ in fm$^{-1}$ is supposed to scale linearly
with $k_F$ in fm$^{-1}$, saturation minima yield the Coester band.

The results with included the 3NF effects are shown in Fig. 5. The N$^2$LO 3NF $V_{123}$
of the chiral effective field theory is first reduced to an effective NN interaction
$V_{12(3)}$ by folding third single-nucleon degrees of freedom.
%\begin{widetext}
\begin{align}
  & \langle \bk_1' \sigma_1'\tau_1',\bk_2' \sigma_2'\tau_2'|V_{12(3)}| \bk_1 \sigma_1\tau_1,
 \bk_2 \sigma_2\tau_2\rangle_A \notag \\
 & \equiv \sum_{\bk_3,\sigma_3 \tau_3} \langle \bk_1' \sigma_1'\tau_1',
 \bk_2' \sigma_2'\tau_2', \bk_3 \sigma_3\tau_3|V_{123}| \bk_1 \sigma_1\tau_1,
 \bk_2 \sigma_2\tau_2, \bk_3 \sigma_3\tau_3\rangle_A,
\end{align}
%\end{widetext}
where $\sigma$ and $\tau$ denote spin and isospin indices.The density-dependent
effective NN force from the N$^2$LO 3NF was first discussed by Holt, Kaiser, and Wesie
\cite{HKW10}. Here, the approximation for the off-diagonal metrix elements in their
paper is not used. The detailed expressions of the partial
wave decomposition are given in Appendices of Ref. \cite{MK13}. When $V_{12(3)}$ is
added to the original NN interaction, some caution is necessary for a statistical factor.
As is explained in Ref. \cite{MK13}, the following prescription for the $G$-matrix
calculations is used:
\begin{equation}
 G_{12}=V_{12}+\frac{1}{3}V_{12(3)}+(V_{12}+\frac{1}{3}V_{12(3)})\frac{Q}{\omega - H}G_{12}.
\end{equation}
The energies in the propagator are calculated by the single-particle energy defined by
\begin{align}
 e_{\bk} &= \langle \bk |t|\bk\rangle + U_G(\bk) \\
 U_G(\bk) &\equiv  \sum_{\bk'} \langle \bk \bk' | G_{12}
  +\frac{1}{6} V_{12(3)}\left(1+\frac{Q}{\omega - H}\right)G_{12} |\bk \bk' \rangle_A .
\end{align}
\begin{figure}[b]
\centering
\includegraphics[width=0.5\textwidth]{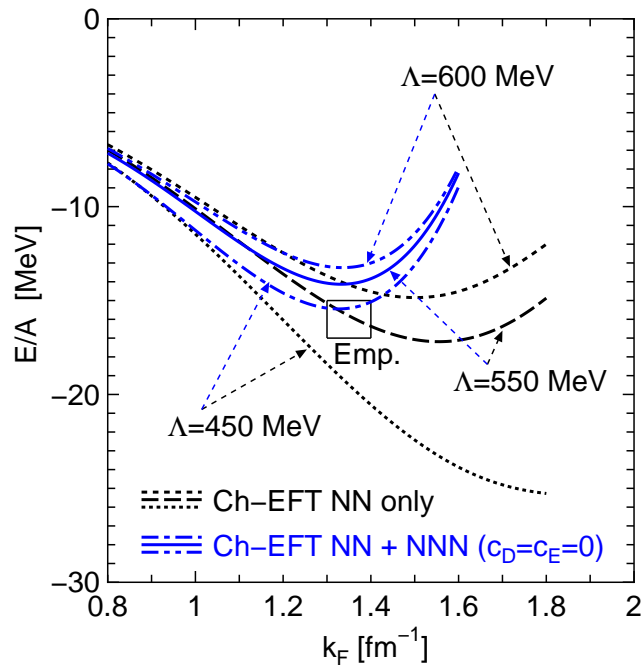}%fig5.eps=sat-cutdep-3nf.eps
\caption{LOBT saturation curves in symmetric nuclear matter with and without the 3NF effects
for three choices of the cutoff energy. The continuous choice is employed for
intermediate spectra. The low-energy constants are set as $c_D=c_E=0$.
}
\end{figure}

\noindent
The total energy per nucleon is evaluated as
\begin{equation}
 E/A = \frac{1}{\rho}\sum_{\bk} \langle \bk |t|\bk\rangle + \frac{1}{2\rho}\sum_{\bk} U_E(\bk),
\end{equation}
where the single-particle potential $U_E(\bk)$ is different from $U_G(\bk)$ given above
and defined by
\begin{equation}
 U_E(\bk) = \sum_{\bk'} \langle \bk \bk' | G_{12} |\bk \bk'\rangle_A.
\end{equation}
The difference between $U_E(\bk)$ and $U_G(\bk)$ may be called a rearrangement
contribution of 3NF origin, which is repulsive and of the order of 5 MeV.

Figure 5 shows that when $V_{12(3)}$ is included the cutoff-energy dependence is
substantially reduced and the saturation curve is improved to match the empirical one.
On the basis of the discussion in the previous subsection, the low-energy constants
of the 3NF contact terms are taken to be $c_D=c_E=0$ as a reference case.
The repulsive contribution arises in the $^1$S$_0$ state. The net contribution of
$^3$P states is also repulsive, although the attractive spin-orbit interaction is
enhanced.

It is helpful to assess the calculated saturation curve by evaluating an incompressibility
$K$, a symmetry energy $S$ and its slope parameter $L$ at the saturation minimum, although
it has to be kept in mind that the calculated curves do not fully reproduce the
empirical saturation minimum. The symmetry energy is estimated by the energies of symmetric
nuclear matter and pure neutron matter, $E_{SNM}(\rho)/A$ and $E_{PNM}(\rho)/A$; that is,
$S=E_{PNM}(\rho_0)/A-E_{SNM}(\rho_0)/A$ where $\rho_0$ is a saturation
density of $E_{SNM}(\rho)/A$, and $L=3\rho_0 \left. \frac{dE_{PNM}(\rho)/A}{d\rho}\right|_{\rho=\rho_0}$.
The saturation curves including the 3NF effects
$V_{12(3)}$ shown in Fig.5 correspond to those values given in Table 2.

\begin{table}[t]
 \caption{Properties of calculated saturation curves in symmetric nuclear matter
 with $V_{12(3)}$ included
 in the three cases of the cutoff energy $\Lambda$. $\rho_0$, $K$, $S$, and $L$
 are a saturation density, an incompressibility, a symmetry energy, and a slope
 parameter, respectively.}
 \setlength{\tabcolsep}{12pt}
 \begin{center}
 \begin{tabular}{cccc} \hline\hline
 $\Lambda$ (MeV) & 450 & 550 & 600 \\ \hline
 $\rho_0$ (fm$^{-3}$) & 0.157 & 0.160 & 0.162 \\
 $E(\rho_0)/A$ (MeV) & $-15.4$ & $-14.1$ & $-13.2$ \\
 $K$ (MeV) & 211 & 209 & 194 \\
 $S$ (MeV) & 34 &  30 & 30  \\
 $L$ (MeV) & 70  & 57 & 53 \\ \hline\hline
 \end{tabular}
 \end{center}
\end{table}

\begin{figure}[b]
\centering
\includegraphics[width=0.9\textwidth]{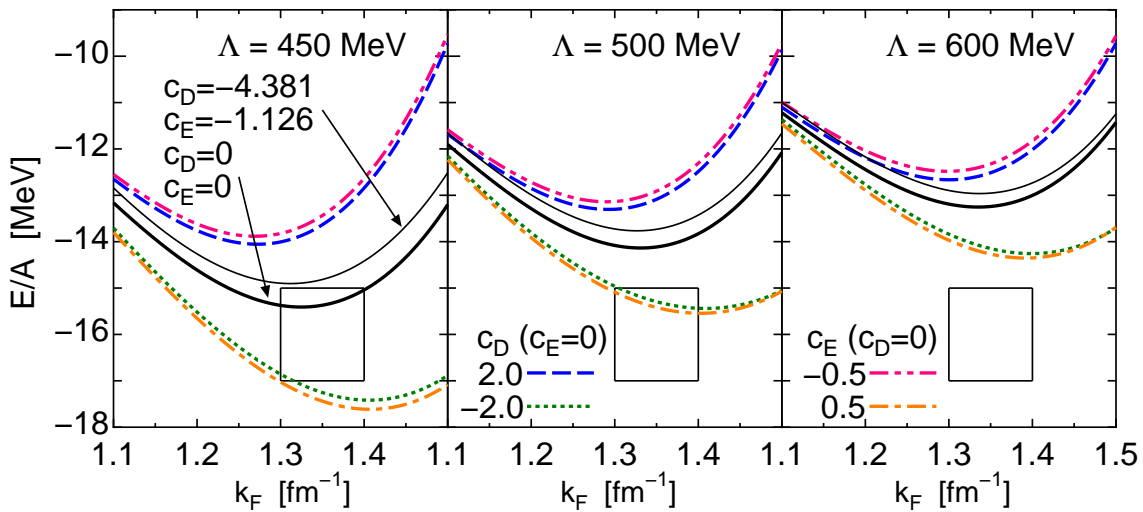}%fig6.eps=cdce-dep3.eps
\caption{Dependence of saturation curves on the 3NF low-energy constants $c_D$ and
$c_E$ around the reference value of $c_D=c_E=0$ for the three choices of the cutoff energy $\Lambda$.
The saturation curves presented in Ref. \cite{MK13} using $c_D=-4.381$ and $c_E=-1.126$ are
also included. 
}
\end{figure}

Let us, now, investigate a range of the changes of the saturation curve due to the
different choice of  $c_D$ and $c_E$. Figure 6 shows four cases of the
saturation curve; that is, $(c_D=2.0,\;c_E=0)$, $(c_D=-2.0,\;c_E=0)$, $(c_D=0,\;c_E=0.5)$,
and $(c_D=0,\;c_E=-0.5)$, compared with the curve of  $(c_D=0,\;c_E=0)$  given in Fig. 5
and that of $(c_D=-4.381,\;c_E=-1.126)$ in Ref. \cite{MK13}. It is seen that if $c_D \approx 4c_E$
is satisfied, contributions from the $c_D$ and $c_E$ terms almost cancel also in the LOBT calculations
and the saturation curve should be close to that of $c_D=c_E=0$. It is noted that the absolute
value of the $c_D$ and $c_E$ term contributions is smaller than the mean-field value, Eqs. (9) and
(10), through the correlation in the $G$-matrix equation.
Another remark is that although the choice of  $c_D=c_E=0$ works well for describing nuclear
saturation properties, a different choice within the relation $c_D\approx 4 c_E$ might be preferred
if a specific spin-isospin character is concerned, because while the $c_D$ term gives rise
to a tensor component, the $c_E$ term does not.

\section{Qualitative explanation of the 3NF effects}
It is meaningful to understand, in a qualitative way, the effects of the 3NF in chiral
effective field theory; that is, to provide the physical interpretation of the 3NF contributions.
The two salient properties are discussed. One is the repulsive effect in the $^1S_0$ channel, which
is basically important to reproduce reasonable saturation properties. The other is the enhancement
of the tensor component in the $^3S_1$ channel. The Pauli blocking plays a central role
in both cases. As far as the influence of the 3NF appears dominantly as the Pauli effect,
the reduction of the 3NF to an effective two-body interaction in the nuclear medium is justified,
because the Pauli effect is operating in two-nucleon processes. It is useful to demonstrate the
consequence of these effects by showing, in subsection 5.3, each $S$ and $T$ channel contribution
in the potential energy.

\subsection{$^1S_0$ channel}
It has been known that if an isobar $\Delta$ excitation in the
two-nucleon scattering process is explicitly considered, the contribution of the diagram depicted in Fig. 7(a)
supplies the (NN)$_{L=0,S=0,J=0}$ state being caused by the tensor component of the pion
exchange, the attraction is similar to that of the tensor correlation in the $^3S_1$
channel. When the $^1S_0$ interaction is embedded in the nuclear medium, the excitation
to the $(\Delta N)$ intermediate state is partly Pauli-blocked, Fig. 7(b).
The suppression of the attraction leads to the repulsive effect. This mechanism has
been studied by various authors. It is also known that the Pauli-blocking effect
for the $\Delta$-excitation can be described by introducing 3-body interaction of
the Fujita-Miyazawa type, Fig. 7(c).

\begin{figure}[b]
\centering
\includegraphics[width=0.5\textwidth]{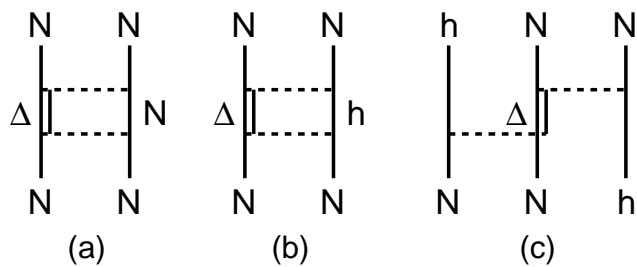}%fig7.eps=pblock1s0.eps
\caption{Diagrams including isobar $\Delta$ excitation: (a) a process to be implicitly
taken into account in the description of NN interaction, (b) "h" standing for an occupied
state, and (c) a 3NF diagram of the Fujita-Miyazawa type.
}
\end{figure}

In the description of NN interaction in chiral effective field theory, the isobar
$\Delta$-excitation does not explicitly appear in most cases. Its effects are
certainly inherent in the coupling constants. The 3NF involving these coupling
constants brings about the relevant Pauli blocking effect.

\begin{figure}[t]
\centering
\includegraphics[width=0.5\textwidth]{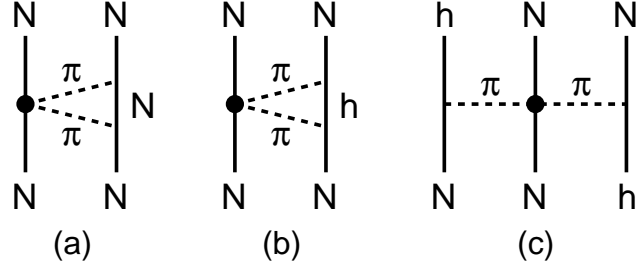}%fig8.eps=pblock3s1.eps
\caption{Two-pion exchange diagrams: (a) a process taken into account in the
description of NN interaction in Ch-EFT, (b) "h" standing for an occupied
state, and (c) a 3NF diagram in Ch-EFT.
}
\end{figure}

\subsection{$^3S_1$ channel}
The mechanism discussed in Sect. 3-1 is irrelevant in the $^3S_1$ channel because of
the isospin. There is, however, another Pauli blocking effect for the tensor component
of the $^3S_1$ NN interaction. The strong tensor force arising from the one-pion
exchange is one of the most salient features of the NN interaction. In the actual
construction of the NN interaction in the one-boson exchange potential (OBEP) model,
the one-pion exchange tensor component has to be weakened to fit the scattering data.
One plausible natural mechanism is the cancellation by the tensor component of
the $\rho$-meson exchange in the opposite sign. In the Ch-EFT, the $\rho$-meson is
irrelevant. However, the reduction of the one-pion exchange tensor component is
found to be given \cite{KBW97} by the two-pion exchange process, Fig. 8(a).
As in the same way for the $\Delta$-excitation, this process is partly Pauli-blocked
in the nuclear medium, Fig. 8(b), and the effect is taken into account by considering
the 3NF, Fig. 8(c).
The suppression of the two-pion exchange process means that the tensor force
is enhanced in the $^3S_1$ channel in the nuclear medium, depending on the
density of nuclear matter, compared with that in the free space. This enhancement
of the tensor component is not expected in the standard OBEP picture.
On the other hand, it has been known that two-pion exchange 3NFs effectively
strengthen the two-nucleon tensor force.

In order to quantify the 3NF effects for the strength of the tensor component
in nuclear matter at the normal density, the bare diagonal matrix elements of the
S-D tensor interaction and those in low-momentum space are shown
in Fig. 9 with and without 3NFs for the 3 cases of the cutoff mass $\Lambda=450$,
$550$, and $600$ MeV, respectively. Figure 9 is for the case of $c_D=c_E=0$. It is seen
that the magnitude of the bare S-D tensor interaction is
enhanced by about 40 \% at the normal density through the 3NF, and this influence
remains in low-momentum space with $\Lambda_{low k}=2$ fm$^{-1}$.
It is interesting that the two-body tensor interactions
with the different cutoff $\Lambda$ become almost identical when they are transformed
into low-momentum space, though the difference is apparent in bare matrix elements.
It is noteworthy that this collapse holds even after 3NF effects are included.
Namely, the equivalent tensor force in low-momentum space is insensitive to the
cutoff scale $\Lambda$.

\begin{figure}[t]
\centering
\includegraphics[width=0.48\textwidth]{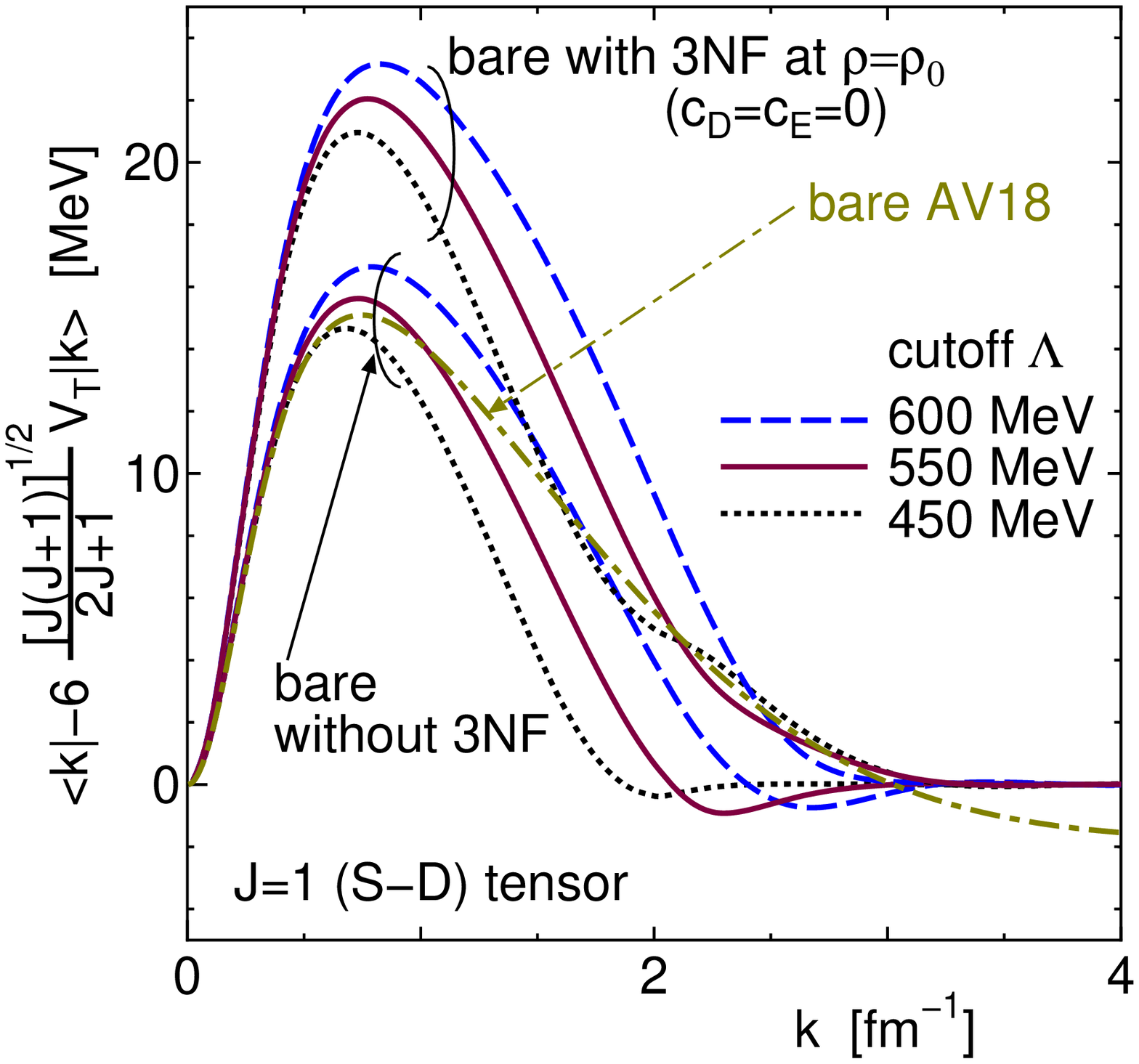}%fig9a.eps=chi-3sd-1t-bare.eps
\hspace{1em}
\includegraphics[width=0.48\textwidth]{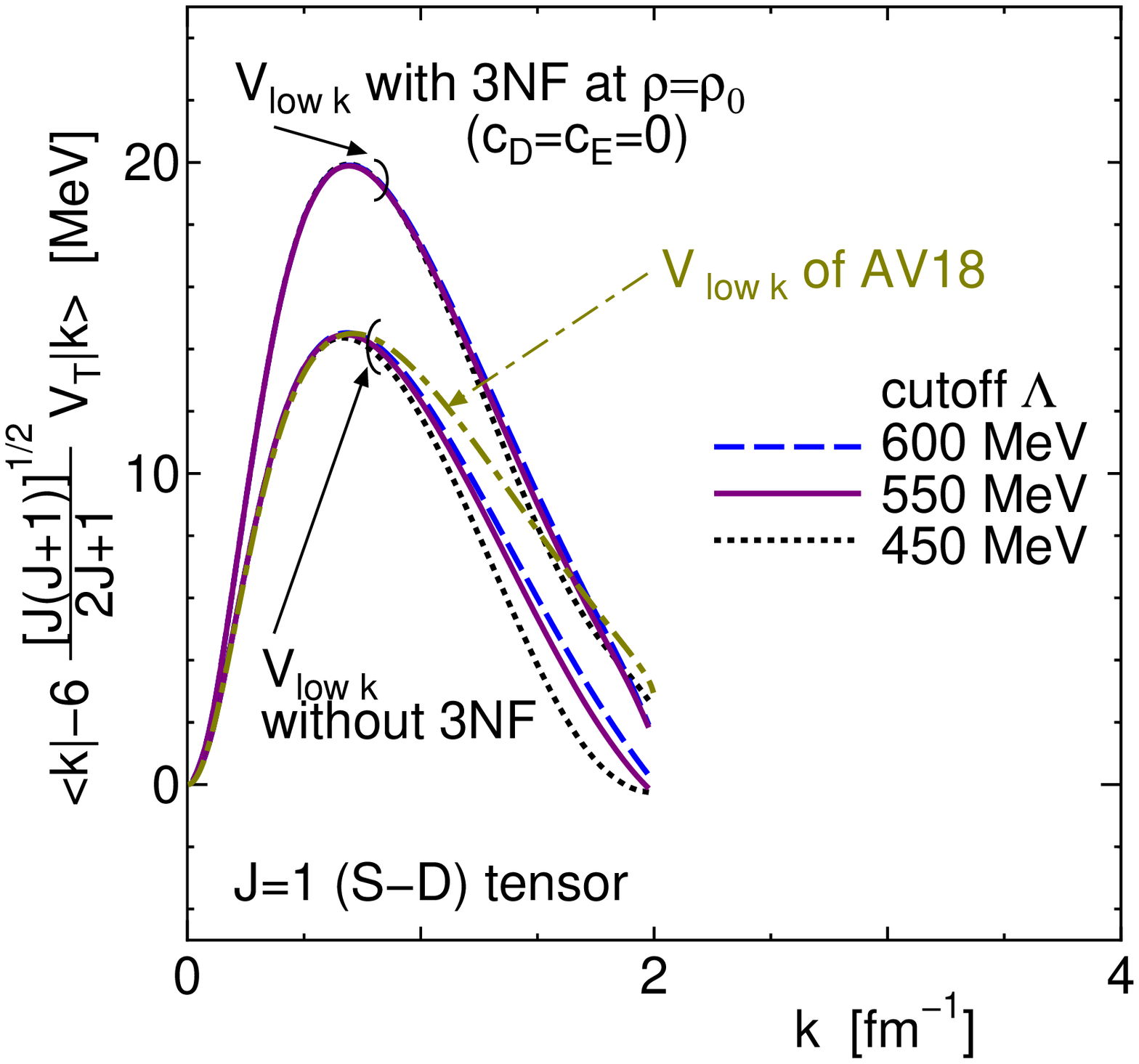}%fig9b.eps=chi-3sd-1t-lowk.eps
\caption{S-D tensor diagonal matrix element in momentum space.
The left panel is for the bare interaction with and without $V_{12(3)}$
for the three cases of the cutoff energy $\Lambda$. The right panel is
for the low-momentum equivalent interaction with $\Lambda_{low k}=2$ fm$^{-1}$.
}
\end{figure}

\begin{figure}[t]
\centering
\includegraphics[width=0.5\textwidth]{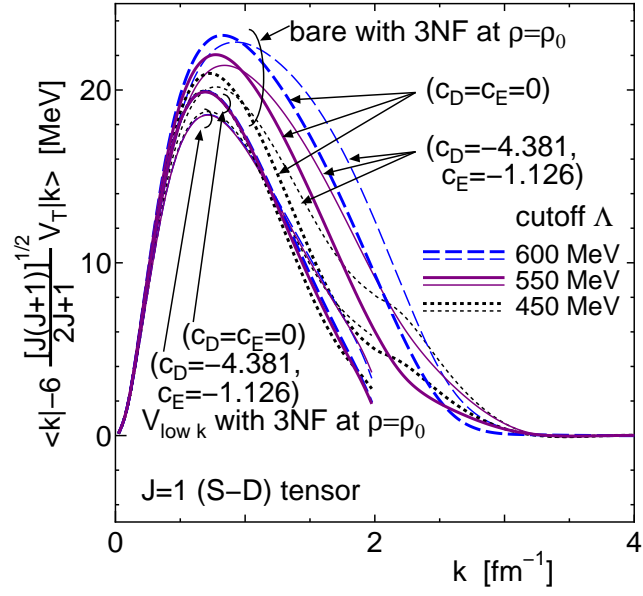}%fig9a.eps=chi-3sd-1t-cdce.eps
\caption{Comparison of bare and low-momentum S-D tensor diagonal matrix elements in momentum space
for the two choices of $c_D$ and $c_E$ parameters: $(c_D=c_E=0)$ and $(c_D=-4.381,\;c_E=-1.126$.
Three cases of the cutoff energy $\Lambda$ are shown. The low-momentum cutoff
is $\Lambda_{low k}=2$ fm$^{-1}$. Note that the curves of the different cutoff energy $\Lambda$ 
for the low-momentum space are virtually indistinguishable.
}
\end{figure}

As noted at the end of the subsection 4.2, the tensor component is brought about through
the $c_D$ term. Figure 10 compares the bare and low-momentum
diagonal S-D matrix elements with $(c_D=c_E=0)$ and those with $(c_D=-4.381,\;c_E=-1.126)$
used in Ref. \cite{MK13}. In the latter case, the enhancement of the magnitude of the S-D tensor
interaction is seen to become smaller to be about 30 \%, which indicates the
quantitative effect of the $c_D$ term of the 3NF to the effective tensor force.

The enhancement of the effective tensor interaction in the nuclear medium due to the 3NF 
might be checked by some adequate experimental observables.
The enhanced tensor interaction of the NN interaction in the nuclear medium
brings about larger attraction in the $^3S_1$ channel through the tensor correlation.
For the nucleon in scattering states, it makes the imaginary part of the optical potential
larger. This effect may be examined by considering nucleon-nucleus and nucleus-nucleus
scattering processes. Promising results are reported in Refs. \cite{MINO15,TOYO15}.

\begin{figure}[b]
\centering
\includegraphics[width=0.4\textwidth]{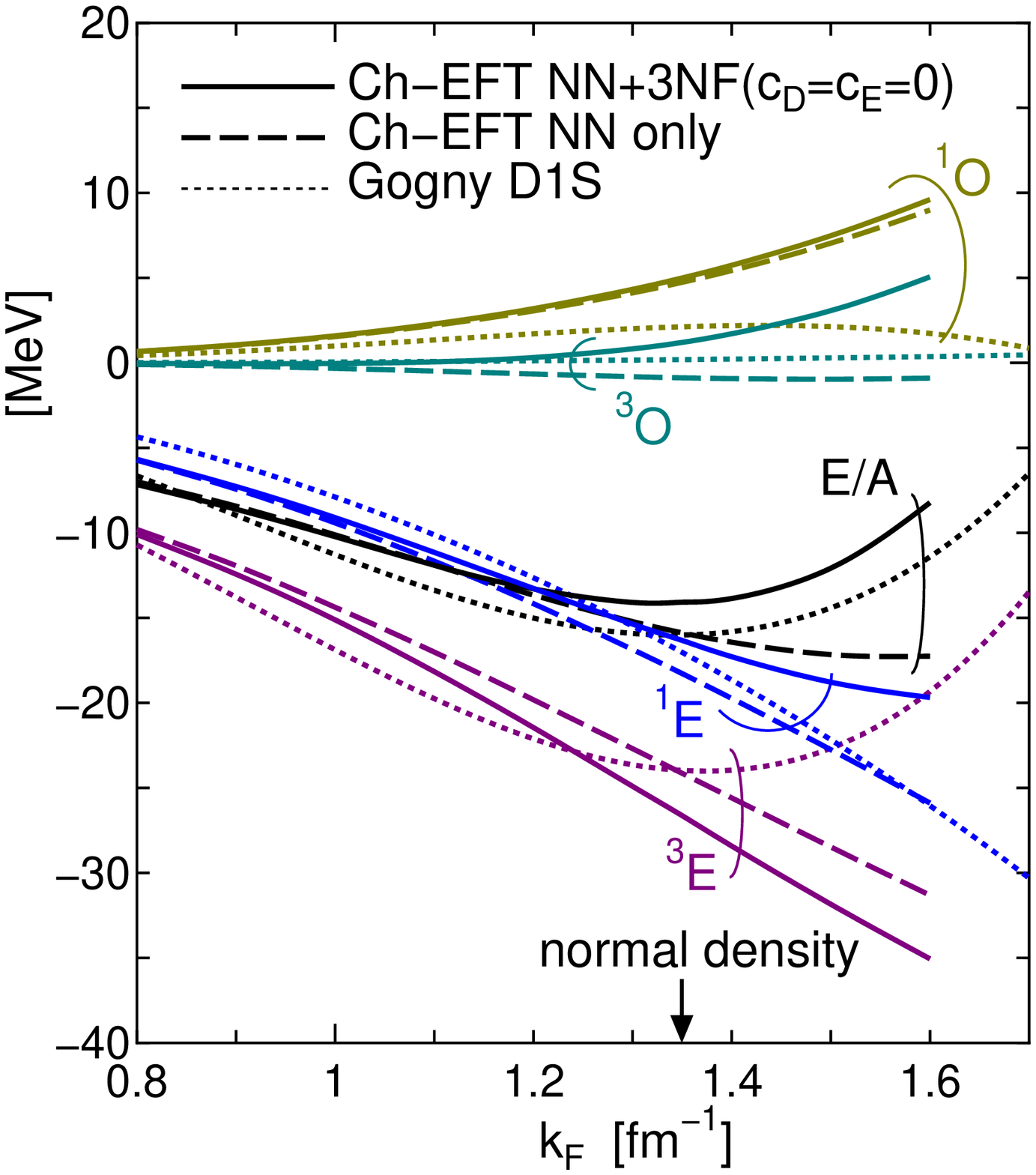}%fig11a.eps=nmpwcD1S-00.eps
\hspace{2em}
\includegraphics[width=0.4\textwidth]{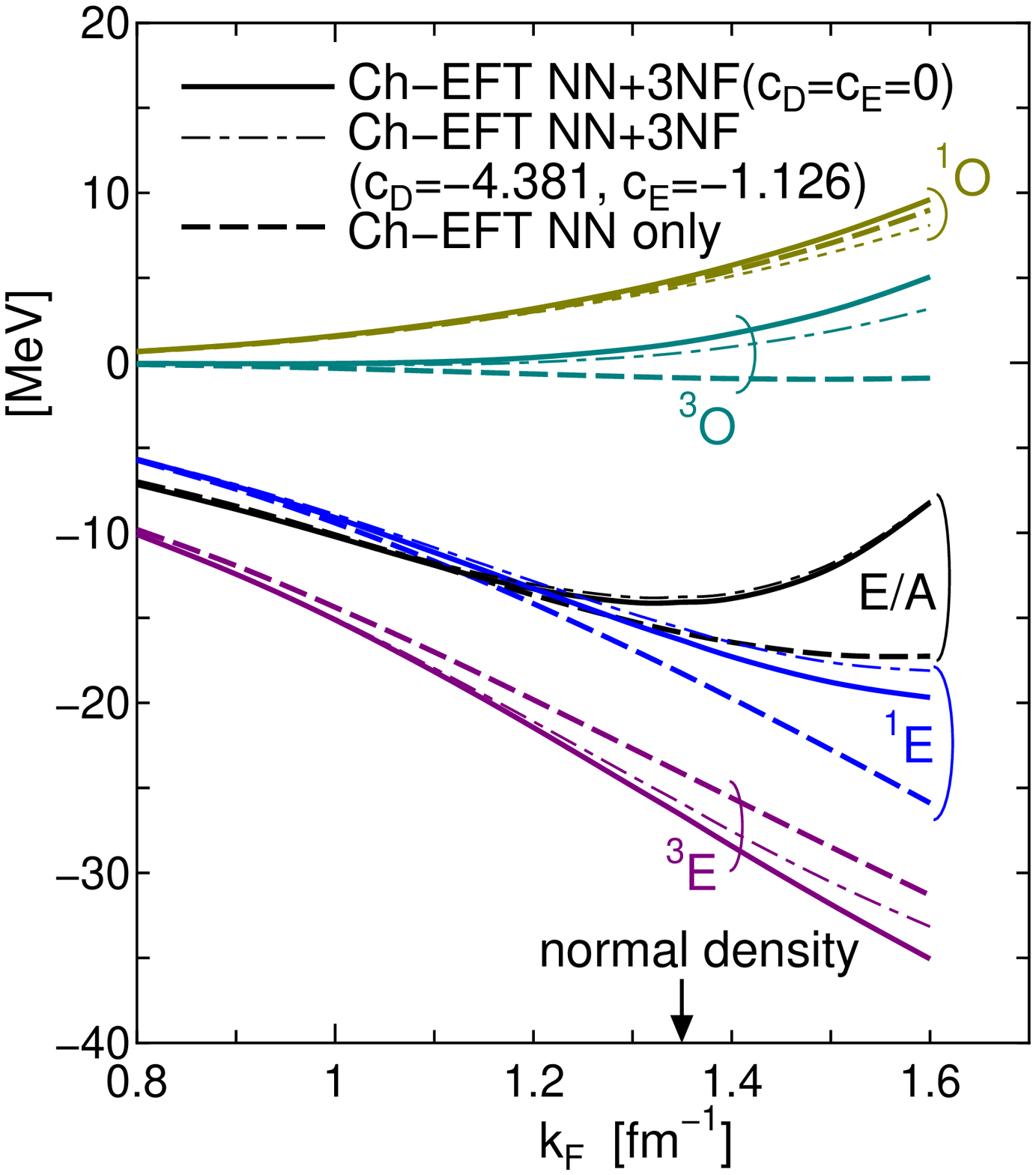}%fig11.eps=nmpwc-cdce.eps
\caption{Decomposition of the potential energy to $^1$O, $^3$E,
$^1E$, and $^3$O contributions in the case of $\lambda=550$ MeV. The left panel is
a comparison of the results with and without 3NF contributions for $c_D=c_E=0$.
The results from the Gogny D1S density-dependent effective interaction \cite{GOGD1S} are
also shown. The right panel includes, in addition, the results with $c_D=-4.381$ and $c_E=-1.126$.
}
\end{figure}

\subsection{Each $S$-$T$ channel contribution of the potential energy}
Summarizing the effects discussed in the preceding subsections, each of $^1$O, $^3$E,
$^1$E, and $^3$O contributions in the potential energy in nuclear matter is given in
Fig. 11 in the case of the cutoff $\Lambda=550$. The corresponding energy obtained
by the Gogny D1S density-dependent effective interaction \cite{GOGD1S}, which has
been commonly used for a mean-field description of finite nuclei, is included.

The difference between the solid and the dashed curves indicates the effect of the 3NF.
The repulsive contribution in the $^3$O channel as well as in the $^1$E channel is
responsible for bringing the saturation minimum at the lower density.
The $^1$O channel is scarcely affected by the 3NF. The largest attractive contribution
in the $^3$E channel increases through the enhanced tensor component, as discussed
in the subsection 3.2. The knowledge of these 3NF contributions may be used to
improve the properties of phenomenological effective interactions. It is interesting
to see that the results of the Ch-EFT interaction with the 3NF effects are similar
to those of the Gogny D1S interaction below the normal density. It is not surprising
that the D1S curves deviates from those of the Ch-EFT at higher density regions,
because the spin-dependent character of the density dependent part is not well
controlled by the data of finite nuclei. While the saturation is assured by the
density-dependent term in the $^3$E channel in the description of D1S, it is
brought about by the repulsion appeared in the $^1$E and $^3$O channels
in the Ch-EFT description.

\section{Summary and remarks}
The saturation mechanism of nuclear matter has been discussed in view of
chiral effective field theory, in which three-nucleon interactions consistent
with the two-nucleon part are systematically defined.
The uncertainties of nuclear matter calculations
in the lowest-order Brueckner theory, which has been known as the Coester band,
were revisited employing modern NN potentials and their low-momentum equivalent
interactions. Then, the results using the chiral N$^3$LO NN and N$^2$LO 3N potentials
with the three choices of cutoff parameters, $\Lambda=450$, 550, and 600 MeV, are
presented. As effective theory, the predictions for empirical quantities are
desirable not to depend on the cutoff scale of the parameterization.
The effective parameters fitted in the two-nucleon interaction level vary
depending on the different cutoff energy, which naturally bear different off-shell
properties, and they are related each other by some unitary transformation.
Therefore, as was shown in Sections 2 and 3, the nuclear matter saturation point
in the LOBT moves on the Coester band depending on the cutoff energy.

It is noteworthy to see that the inclusion of the 3NF changes the situation and
the nuclear saturation is reasonably well reproduced after the 3NF effects are
incorporated. The 3NFs are treated by reducing them to effective
two-body interactions by folding the third nucleon degrees of freedom.
As explained in the discussion of Sec. 5, this approximation is justified by
the fact that the main effects of the 3NF can be regarded as
resulting from the inclusion of the Pauli effects in two-pion exchange
processes of the two-nucleon interaction. It is to be noted
that the contributions of the specific 3NF processes bearing new
parameters $c_D$ and $c_E$, turn out to give minor effects, as far as the relation
$c_D \approx 4 c_E$ is satisfied. The prescription of $c_D\approx 4c_E$ is
favorable for explaining nuclear saturation properties in the LOBT. Thus,
calculations using $c_D=c_E=0$ are mainly presented as a reference case.
Then, the dependence of the calculated result on the choice of $c_D$ and $c_E$
is shown. If a specific component of the effective NN interaction in the
nuclear medium is interested, fine tuning of $c_D$ and $c_E$ may be applied.
%which has been actually found in various few-body
%calculations and perturbation calculations in nuclear matter.
Except for the reasonable choice of $c_D$ and $c_E$, being in a natural size,
no adjustable parameter is introduced in this paper.

The repulsive contribution in the $^1S_0$ state is understood as the effect
of the suppression of the isobar $\Delta$ excitation in the nuclear medium,
which has been known a long time, although there is no explicit $\Delta$
in the present N$^3$LO interaction. In the $^3S_1$ channel, the isobar $\Delta$ is
irrelevant. In this case, the important effect appears through the tensor component.
The one-pion exchange has a strong tensor component. To match the experiment,
the strength is needed to be reduced. A physical and natural cut is
provided, with a description of a OBEP model, by the $\rho$-meson exchange which has
a tensor component in the opposite
sign. In chiral effective theory, a two-pion exchange process replace the role
of the $\rho$-meson exchange, and the two-pion exchange is hindered in the nuclear medium
to partly restore the one-pion exchange tensor component.

The significant role of the 3NF in realizing correct nuclear saturation properties
indicates that other baryon degrees of freedom than protons and neutrons,
which are eliminated in constructing potential description, are important.
The 3NF is analogous to the appearance of induced many-body interactions
in a restricted space of nuclear many-body theory. Though the reason of
justifying the reduction of the 3NFs to an effective two-nucleon interaction in the medium
is explained, it is certain that a full treatment of the 3NF is necessary.
Concerning this problem, it is noted as a final remark that the wound
integral $\kappa$ of the LOBT calculation with the Ch-EFT potential is
somewhat large, in particular when the 3NF effects are included,
$\kappa \simeq 0.25$, compared to that of other modern NN potentials which
is around $\kappa \simeq 0.15$. This suggests that the convergence of
the Brueckner-Bethe-Goldstone hole-line expansion is slower.
It is therefore important in the future to quantitatively estimate more than three-nucleon
correlation energies in the nuclear medium using the Ch-EFT interaction.
%In fact, calculations in finite nuclei using the unitary model operator
%approach \cite{SOK87} show that three-nucleon correlation energies are
%considerably larger than those of other modern NN potentials.
%These details will be reported in a separate paper \cite {KO15}.

\section*{Acknowledgment}
This work is supported by the Japan Society for the Promotion of Science (JSPS)
KAKENHI Grant No. 25400266.


\begin{thebibliography}{99}
\bibitem{YUK35} H. Yukawa, Proc. Physico-Mathematical Society of Japan, \textbf{17},
48 (1935).
\bibitem{BLM54} K.A. Brueckner, C.A. Levinson, and H.M. Mahmoud, Phys. Rev.
\textbf{95}, 217 (1954).
\bibitem{GOL57} J. Goldstone, Proc. Roy. Soc. \textbf{A239}, 267 (1957).
\bibitem{DAY67} B.D. Day, Rev. Mod. Phys. \textbf{39}, 719 (1967).
\bibitem{BET71} H.A. Bethe, Ann. Rev. Nucl. Sci.\textbf{21}, 93 (1971).
\bibitem{DAY78} B.D. Day, Rev. Mod. Phys. \textbf{50}, 495 (1978).
\bibitem{LSZ06} Z.H. Li, U. Lombardo, H.-J. Schulze, W. Zuo, L.W. Chen, and H.R. Ma,
Phys. Rev. \textbf{C74}, 047304 (2006).
\bibitem{SBGL98} H.Q. Song, M. Baldo, G. Giansiracusa, and U. Lombardo, Phys. Rev. Lett.
\textbf{81}, 1584 (1998).
\bibitem{PW79} V.R. Pandharipande and R.B. Wiringa, Rev. Mod. Phys. \textbf{51}, 821 (1979).
\bibitem{BB12} M. Baldo and G.F. Burgio, Rep. Prog. Phys. \textbf{75}, 026301 (2012).
\bibitem{FM57} J. Fujita and H. Miyazawa, Prog. Theor. Phys \textbf{17}, 366 (1957).
\bibitem{BM90} R. Brockmann and R. Machleidt, Phys. Rev. C \textbf{42}, 1965 (1990).
\bibitem{BG69} G.E. Brown and A.M. Green, Nucl. Phys. A \textbf{137}, 1 (1969).
\bibitem{LNR71} B.A. Loiseau, Y. Nogami, and C.K. Ross, Nucl. Phys. A \textbf{165}, 601 (1971);
Erratum A \textbf{176}, 665 (1971).
\bibitem{KAT74} T. Kasahara, Y. Akaishi, and H. Tanaka, Prog. Theor. Phys. Suppl. \textbf{56},
96 (1974).
\bibitem{FP81} B. Friedman and V.R. Pandharipande, Nucl. Phys. A \textbf{361}, 361 (1981).
\bibitem{EHM09} E. Epelbaum, H.-W. Hammer, and U.-G. Meisner, Rev. Mod.
Phys. \textbf{81}, 1773 (2009).
\bibitem{ME11} R. Machleidt and D.R. Entem, Phys. Rep. \textbf{503}, 1 (2011).
\bibitem{BSFN05} S.K. Bogner, A. Schwenk, R.J. Furnstahl, and A. Nogga,
Nucl. Phys. A \textbf{763}, 59 (2005).
\bibitem{HS10} K. Hebeler and A. Schwenk, Phys. Rev. C \textbf{82}, 014314 (2010).
\bibitem{HEB11} K. Hebeler, S.K. Bogner, R.J. Furnstahl, A. Nogga,
and A. Schwenk, Phys. Rev. C \textbf{83}, 031301(R) (2011).
\bibitem{CORA14} L. Coraggio, J.W. Holt, N. Itaco, R. Machleidt, L.E. Marcucci
and F. Sammarruca, Phys. Rev. C \textbf{89},044321 (2014).
\bibitem{MK12} M. Kohno, Phys. Rev. C \textbf{86}, 061301(R) (2012).
\bibitem{HKW10} J.W. Holt, N. Kaiser, and W. Weise, Phys. Rev. C {\bf 81}, 024002 (2010).
\bibitem{MK13} M. Kohno, Phys. Rev. C \textbf{88}, 064005 (2013).
\bibitem{SCC12} F. Sammarruca, B. Chen, L. Coraggio, N. Itaco, and R. Machleidt,
Phys. Rev. C \textbf{86}, 054317 (2012).
\bibitem{SCC15} F. Sammarruca, L. Coraggio, J. W. Holt, N. Itaco, R. Machleidt, and
L. E. Marcucci, Phys. Rev. C \textbf{91}, 054311 (2015).
\bibitem{CARB13} A. Carbone, A. Cipollone, C. Barbieri, A. Rios, and A. Polls,
Phys. Rev. C \textbf{88}, 054326 (2013).
\bibitem{CARB14} A. Carbone, A. Rios, and A. Polls,
Phys. Rev. C \textbf{90}, 054322 (2014).
\bibitem{BKS03} S. K. Bogner, T. T. S. Kuo, and A. Schwenk, Phys. Rep. 386, 1 (2003).
\bibitem{COE70}  F. Coester, S. Cohen, B. Day, and C.M. Vincent, Phys. Rev. C \textbf{1},
769 (1970).
\bibitem{SMC99} E. Schiller, H. M\"{u}ther, and P. Czerski, Phys. Rev. C \textbf{59},
2934 (1999);
\textbf{60}, 059901 (1999).
\bibitem{SOKN00} K. Suzuki, R. Okamoto, M. Kohno, and S. Nagata,
Nucl. Phys. A \textbf{665}, 92 (2000).
\bibitem{AV18} R.B. Wiringa, V.G.J.Stoks, and R. Schiavilla,
Phys. Rev. C \textbf{51}, 38 (1995).
\bibitem{NSC} T. A. Rijken, V. G. J. Stoks, and Y. Yamamoto, Phys. Rev. C \textbf{59},
21 (1999).
\bibitem{CDB} R. Machleidt, Phys. Rev. C \textbf{63}, 024001 (2001).
\bibitem{FSS2} Y. Fujiwara,Y. Suzuki, and C. Nakamoto, Prog. Part. Nucl. Phys.
\textbf{58}, 439 (2007).
\bibitem{SL80} K. Suzuki and S. Y. Lee, Prog. Theor. Phys. 64, 2091 (1980).
\bibitem{LS80} S. Y. Lee and K. Suzuki, Phys. Lett. B91, 173 (1980).
\bibitem{KMMS03} J. Kuckei, F. Montani, H. M\"{u}ther, and A. Sedrakian, Nucl. Phys.
\textbf{A723}, 32 (2003).
\bibitem{JNF11} E.D. Jurgenson, P. Navratil, and R.J. Furnstahl, Phys. Rev. \textbf{C83},
034301 (2011).
\bibitem{SOK87} K. Suzuki, R. Okamoto, and Kumagai, Phys. Rev. \textbf{C36}, 804 (1987).
\bibitem{FOS09} S. Fujii, R. Okamoto, and K. Suzuki, Phys. Rev. Lett. \textbf{103},
182501 (2009).
\bibitem{EGM05} E. Epelbaum, W. G\"{o}ckle, and U.-G. Mei{\ss}ner,
Nucl. Phys. A \textbf{747}, 362 (2005).
\bibitem{NGV07} P. Navr\'{a}til, V.G. Gueorguiev, J.P. Vary, W.E. Ormand, and A. Nogga,
Phys. Rev. Lett. \textbf{99}, 042501 (2007).
%\bibitem{NBS04} A. Nogga, S.K. Bogner, and A. Schwenk,
%Phys. Rev. C \textbf{70}, 061002(R) (2004).
\bibitem{KBW97} N. Kaiser, R. Brockmann, and W. Weise,
Nucl. Phys. A \textbf{625}, 758 (1997).
\bibitem{TOYO15} M. Toyokawa, K. Minomo, M. Kohno, and M. Yahiro,
J. Phys. G: Nucl. Part. Phys. 42, 025104 (2015).
\bibitem{MINO15} K. Minomo, M. Toyokawa, M. Kohno, and M. Yahiro,
Phys. Rev. \textbf{C90}, 051601(R) (2014).
\bibitem{GOGD1S}J. F. Berger, M. Girod, and D. Gogny,
Comput. Phys. Commun. \textbf{63}, 365 (1991).
%\bibitem{KO15} M. Kohno and R. Okamoto, in preparation.
%-----------------
\end{thebibliography}
\end{document}